\documentclass[a4paper,11pt]{article}
\pdfoutput=1 % if your are submitting a pdflatex (i.e. if you have
\usepackage{jinstpub} % for details on the use of the package, please
\usepackage{amsmath}

\usepackage{xcolor}

\usepackage{subcaption}

\title{\boldmath Simulation of muon-induced neutral particle background for a shallow depth Iron Calorimeter detector}

\author[a,b,1]{Neha Panchal,\note{Corresponding author.}}
\author[b]{G. Majumder}
\author {and}
\author[b]{V.M. Datar}
\affiliation[a]{Homi Bhabha National Institute, \\ Anushaktinagar, Mumbai-400094, India}
\affiliation[b]{Tata Institute of Fundamental Research, \\Homi Bhabha Road, \\Colaba, Mumbai-400005, India}
\emailAdd{neha.dl0525@gmail.com}

\abstract{The Iron Calorimeter (ICAL) detector at the India based Neutrino Observatory (INO) is planned to be set up in an underground cavern with a rock overburden of more than 1\,km. This overburden reduces the cosmic muon flux by a factor of 10$^{6}$ with respect to that at the sea level. In this paper, we examine the possibility of a 100\,m shallow depth ICAL (SICAL) detector. The cosmic muons would have to be detected in a veto detector with an efficiency of 99.99\% in order to have the same level of muon background leaking undetected through the veto detector as at the  1\,km depth underground site. However, an additional background arises from interactions of cosmic muons with the rock. Since the neutral particles produced in such interactions can pass through the veto detector without any interaction, they can possibly mimic neutrino events in the ICAL. In this paper, the results of a GEANT4 based simulation study to estimate the background signals due to muon induced interactions with the rock for the SICAL is presented.}


\arxivnumber{1234.56789} % only if you have one

\begin{document}
\maketitle
\flushbottom

\section{Introduction}
\label{sec:intro}

The proposed Iron Calorimeter (ICAL) at the India-based Neutrino Observatory (INO) will consist of three 17\,kton modules each having 150 layers of iron interleaved with Resistive Plate Chamber (RPC) detectors \citep{akumar}. The major goal of the ICAL is to determine the mass hierarchy in the neutrino sector \citep{ghosh2013}. INO will be situated at Pottipuram in Bodi West hills of Theni, India under a mountain cover of about 1\,km from all the sides.  The rock overburden above the ICAL significantly reduces the background arising due to cosmic ray muons (by a factor of 10$^{6}$). A background free, or a very low background, environment is indispensable for searching rare physics processes. To reduce the background level one needs to first understand all the sources of background and then design a technique to reduce it. One such technique is the use of a cosmic muon veto detector (CMVD), the use of which, in the context of shallow depth the ICAL has been discussed in \citep{npanchal}. The reduction in cosmic ray muon flux at a depth of 100\,m is about 10$^{2}$. Therefore, the veto efficiency required to achieve the background level of the Theni site is 99.99\%, a goal which appears to be feasible \citep{npanchal}. At shallow depths, the most significant background is due to cosmic ray muons which can be vetoed with a high veto efficiency. However, the secondaries generated due to muon-nucleus interaction with the rock material can be a serious concern which needs to be carefully investigated. High energy cosmic muons can undergo an inelastic interaction with the nucleus or could be absorbed. The secondary neutral particles produced in such interactions could escape detection in the veto detector and give rise to events in the ICAL that could mimic a neutrino interaction. The SICAL is a worthwhile proposition if the number of such false positives are reduced to a level that is significantly lower than the true positives (i.e. genuine neutrino events). It may not, therefore, be out of place to mention some of possible advantages of the SICAL detector viz.
\\
(1) a much larger choice of sites,\\
(2) much larger caverns,\\
(3) monitoring of RPCs using the much larger muon flux at 100\,m depth as compared to that at 1\,km depth,\\
(4) if shown to be feasible, using muon spin rotation to get additional information on the internal magnetic field \citep{muSR} and\\
(5) enhancing the sensitivity for exotic searches including probing the origin of the anomalous KGF events \citep{dash1} and cosmogenic magnetic monopoles \citep{dash2} using the CMVD and the ICAL.

In this paper, we present results of a Monte Carlo study to estimate the contribution of this background and compare it with the event rate expected from atmospheric neutrinos in the SICAL detector. We first present the details of the simulation framework in Section 2. The outcome of the simulation by propagating the neutrals through the ICAL detector using the INO-ICAL simulation code~\cite{ical_update} is discussed in the Section 3. In Section 4, we discuss the estimation of the false positive event rate at the SICAL. The summary of the major results of the simulation is presented in Section 5.
%The experimental requirements for searching for rare physics processes requires a background free, or a very low background, environment.

\section{Simulation framework}

\begin{figure}[ht]
  \centering
    \includegraphics[width=100mm]{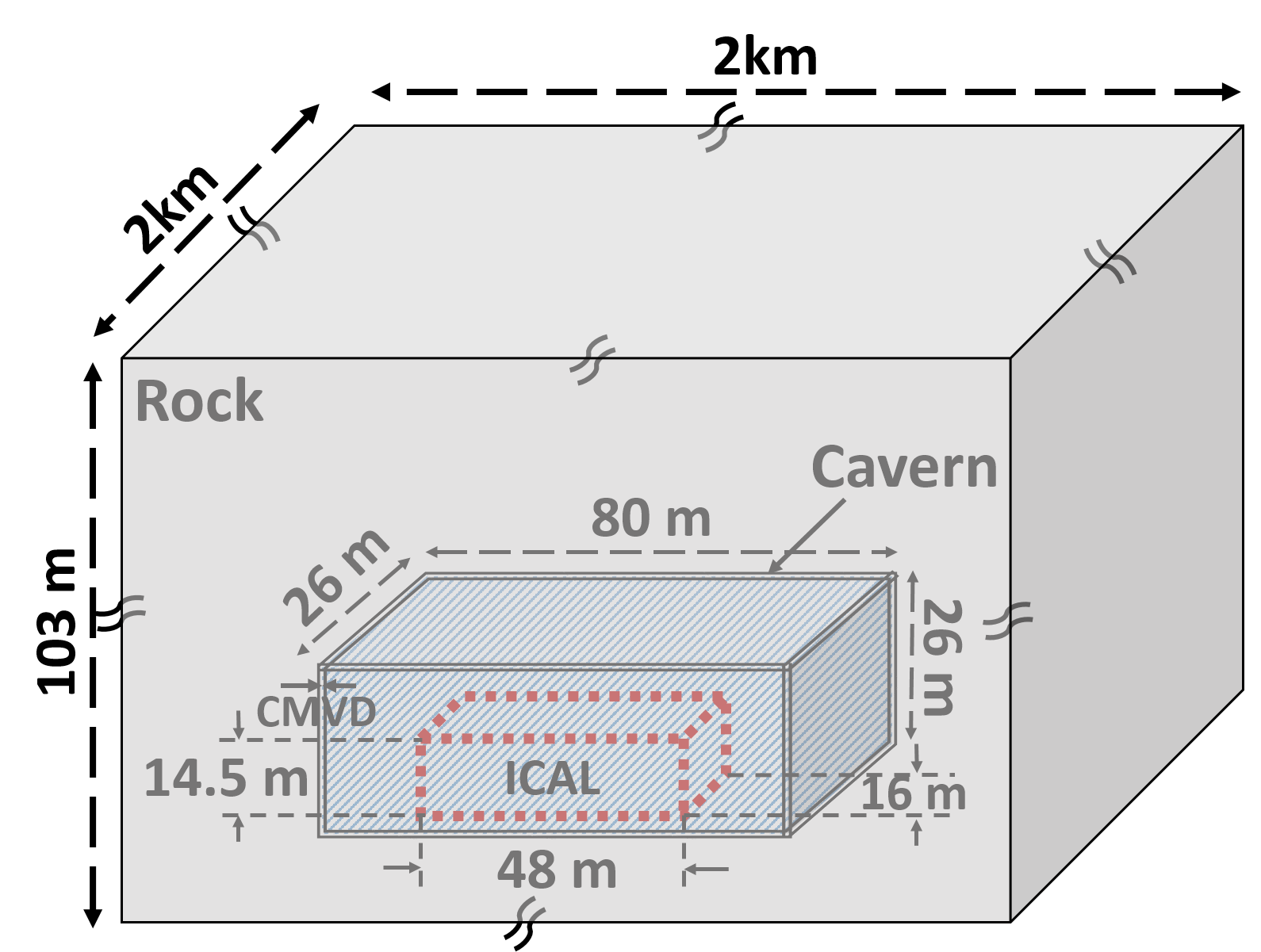}

 \caption{A schematic of the SICAL detector at a depth of 103\,m along with the CMVD}
 
  \label{fig:SCICAL}
\end{figure}

A schematic of the SICAL detector at a depth of 103\,m with the CMVD is shown in Fig.~\ref{fig:SCICAL}. The cavern $\rm(80\,m\,\times\,26\,m\,\times\,26\,m )$ is surrounded by a rock coverage of 2\,km from both the horizons and the CMVD, made up of 3\,cm thick scintillators, is placed up against the 4 walls and the ceiling of the cavern.
%%%%%%%%%%%%%%%%%%%%%%%%%%%%%%%%%%%%%%%%%%%%%%%%%%%%%%%%%%%%
%%%%%%%%%%%%%%%%%%%%%%%%%%%%%%%%%%%%%%%%%%%%%%%%%%%%%%%%%%%
A full simulation of the neutrino-like events in the ICAL involves propagating the cosmic muons at the surface in the intervening rock corresponding to the chosen location. This would include keeping track of the secondary particles produced in muon-nucleus interactions anywhere along their path towards the ICAL detector at a specified depth. We performed the simulation for a depth of 103\,m and the results are presented in this paper. The propagation of low energy muons and secondary hadrons, produced in high energy muon interactions in the upper part of the 103\,m rock overburden, increases the computation time although these particles will not survive the remaining rock thickness. Therefore, to decrease the computation time, the simulation was done in two parts with the corresponding geometry using GEANT4 \citep{GEANT4}. In the first part, the cosmic muon flux is generated at a depth of 100\,m. Using this flux, muon-nuclear interactions are simulated in the last 3\,m of rock and the neutrals produced are propagated to the ICAL detector in the second part. 

\begin{figure}[ht]
  \centering
  \begin{tabular}{ccc}
    \includegraphics[width=55mm]{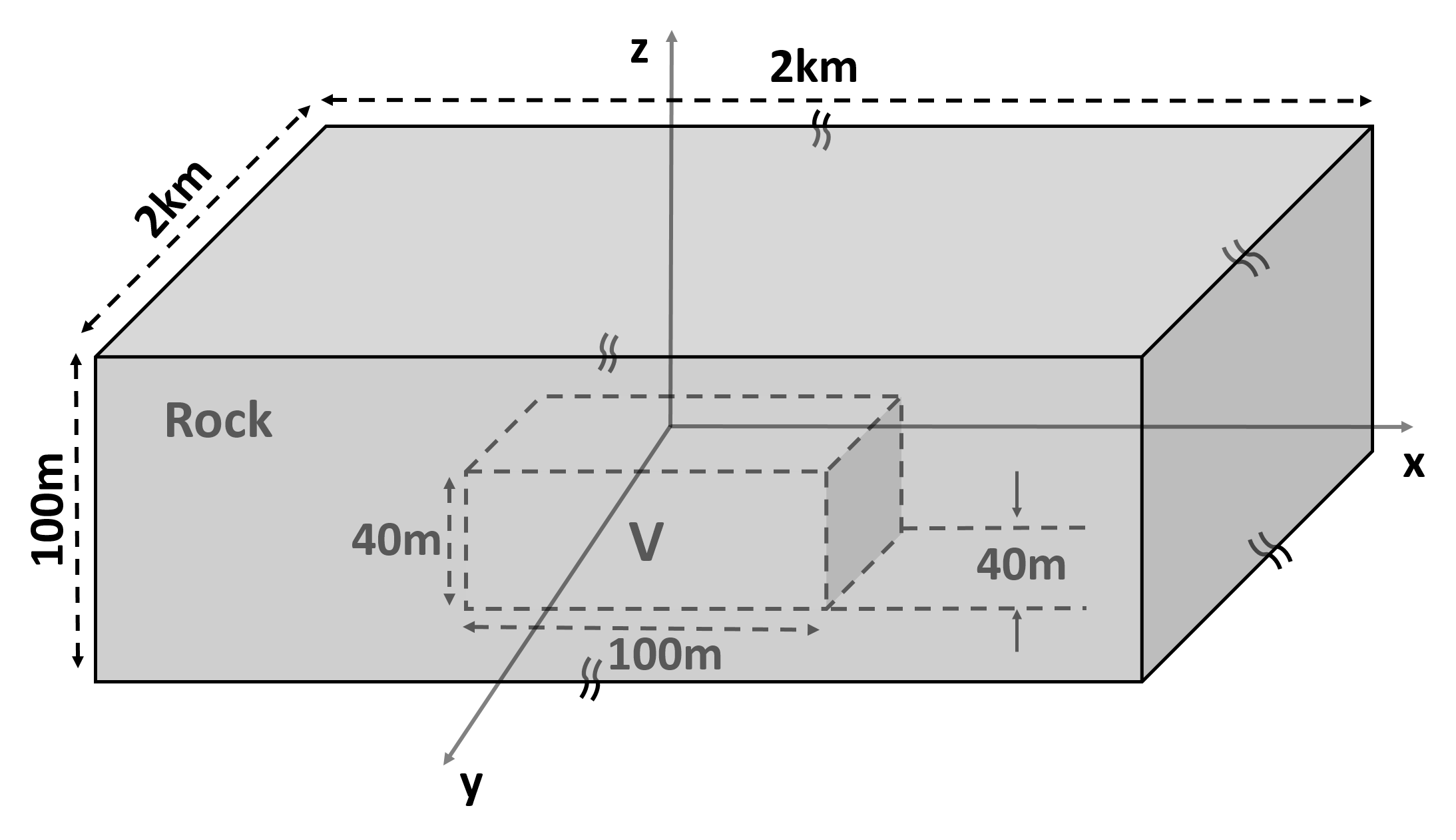} &
    \includegraphics[width=45mm]{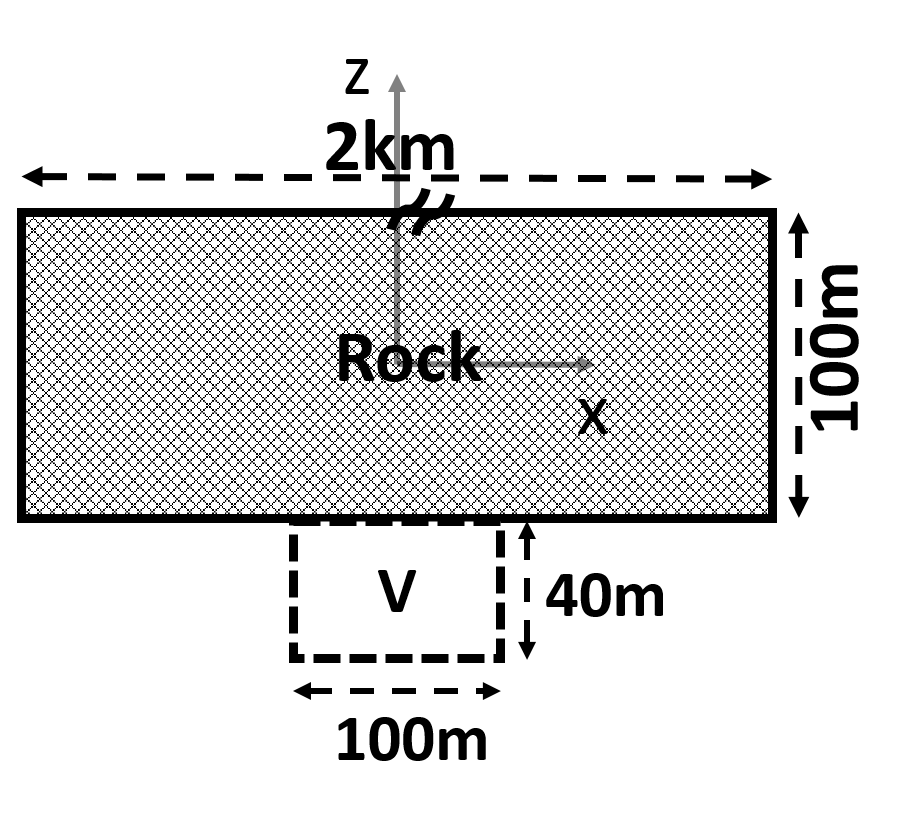} &
    \includegraphics[width=45mm]{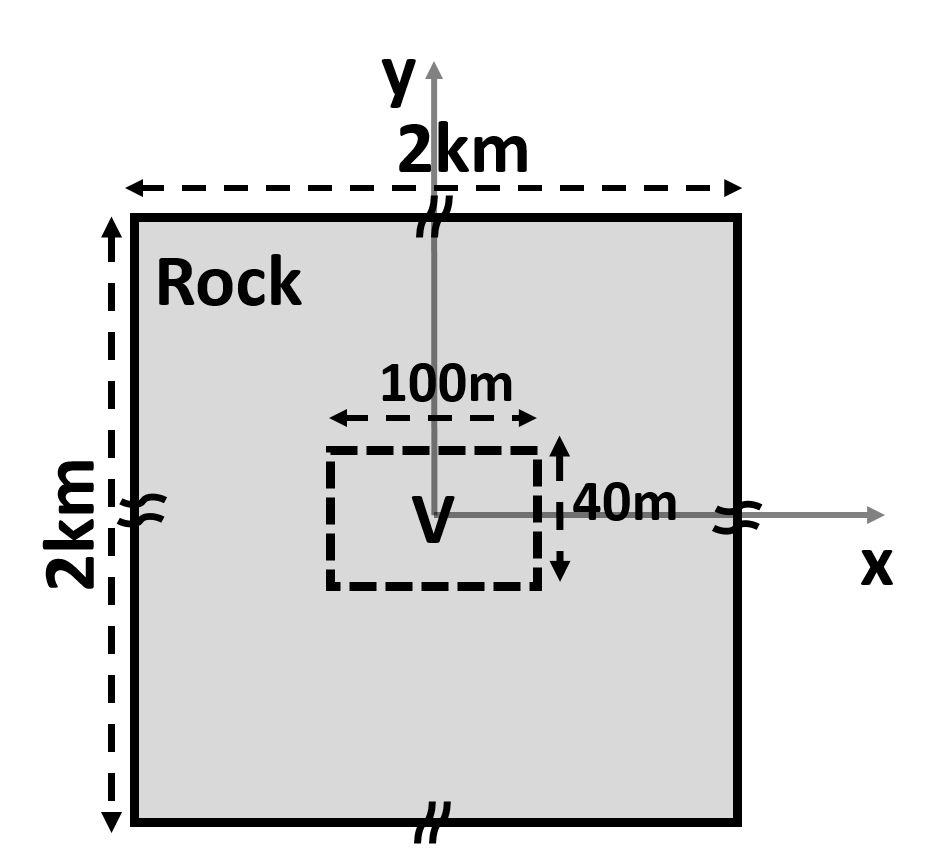}
  \end{tabular}
 \caption{A schematic representation for the geometry used for the first part of the simulation. 3D cartoon picture (\textit{left}), 2D side view (\textit{center}) and top view (\textit{right}).}
 
  \label{fig:schem_part1}
\end{figure}

\begin{figure}[ht]
  \centering
  \begin{tabular}{ccc}
    \includegraphics[width=45mm]{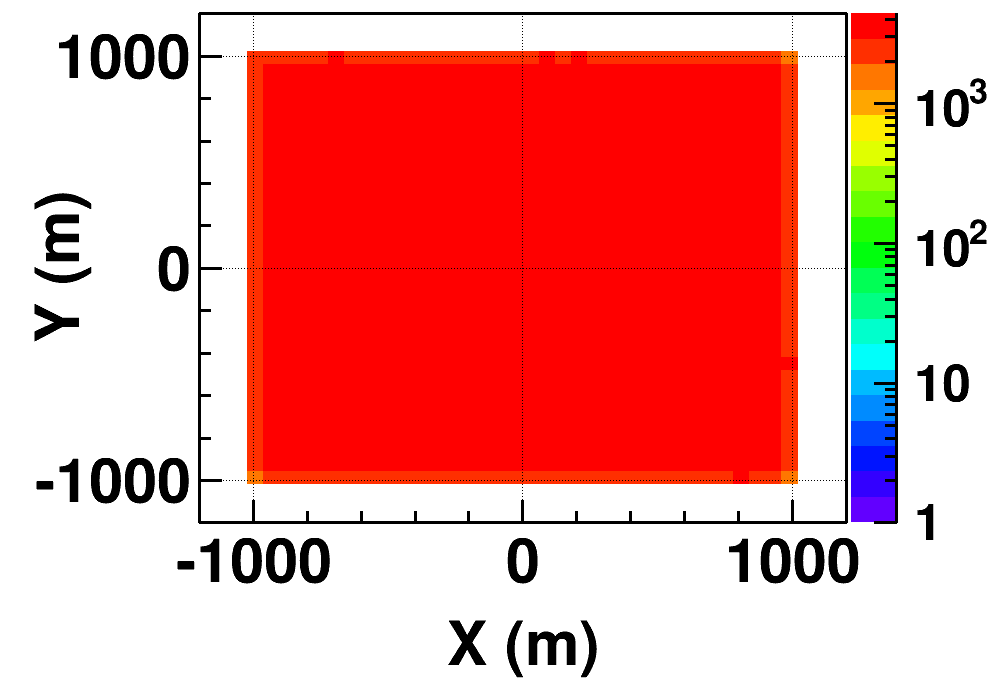} &
    \includegraphics[width=45mm]{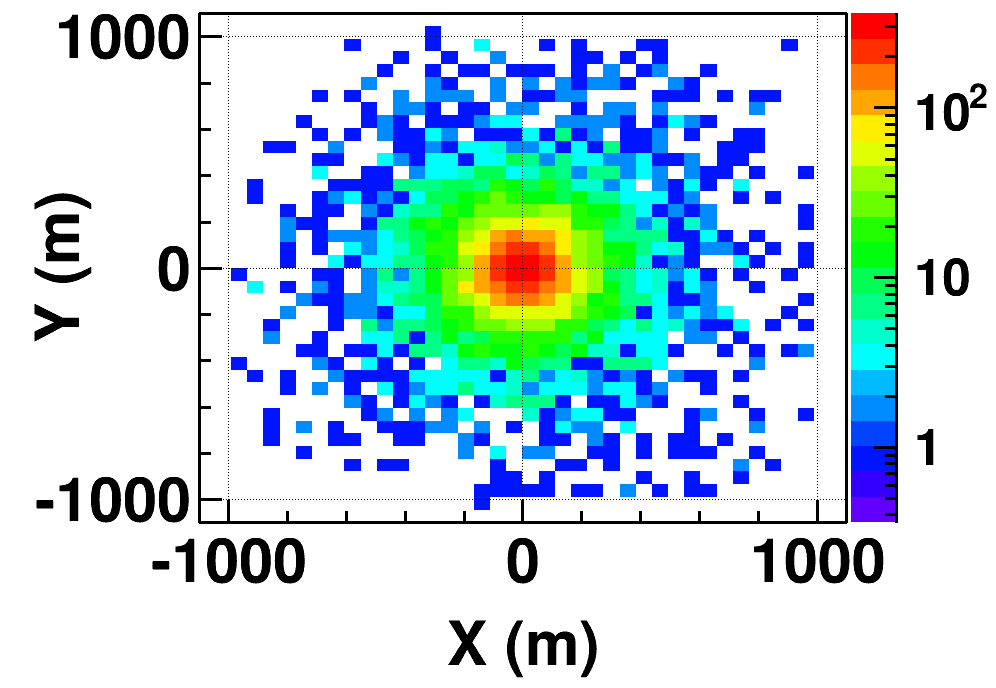}&
    \includegraphics[width=45mm]{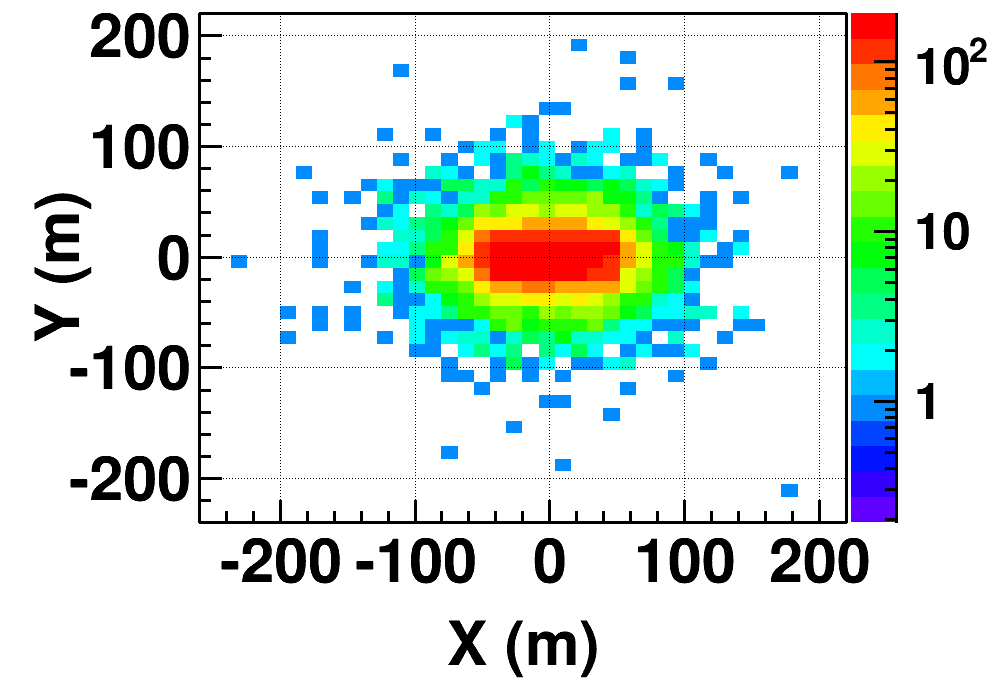}\\
  \end{tabular}
 \caption{The position distribution, (x,y) of events at the rock surface for $N_0$ (\textit{left}), $N_1$ (\textit{center}) and $N_2$ (\textit{right}).}
 
  \label{fig:xy_pos_N0N1N2}
\end{figure}

\begin{figure}[ht]
  \centering
  \begin{tabular}{ccc}
    \includegraphics[width=45mm]{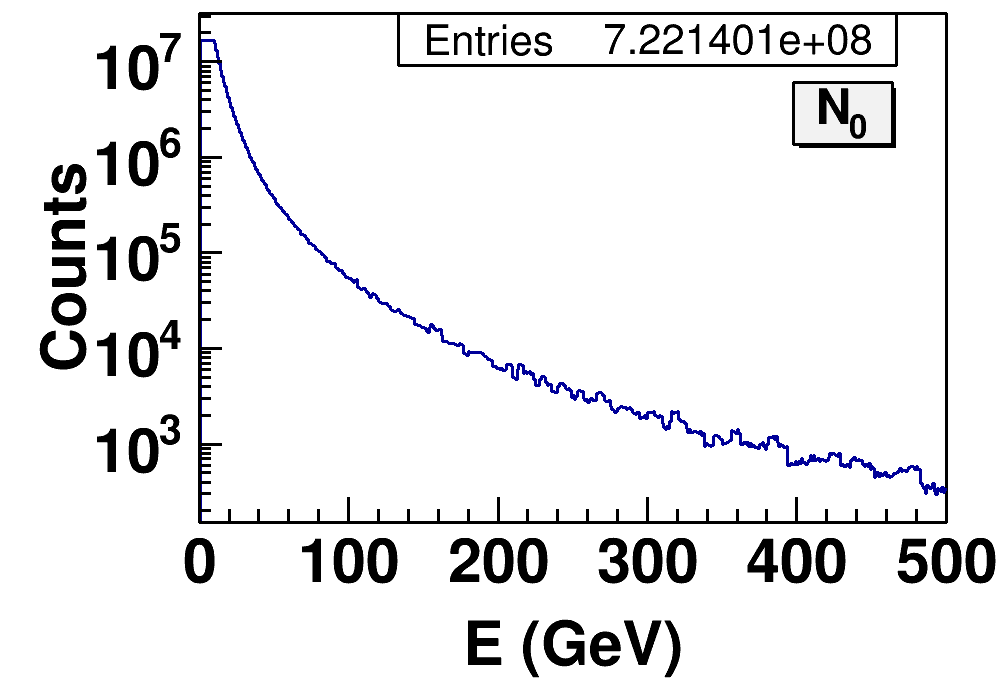} &
    \includegraphics[width=45mm]{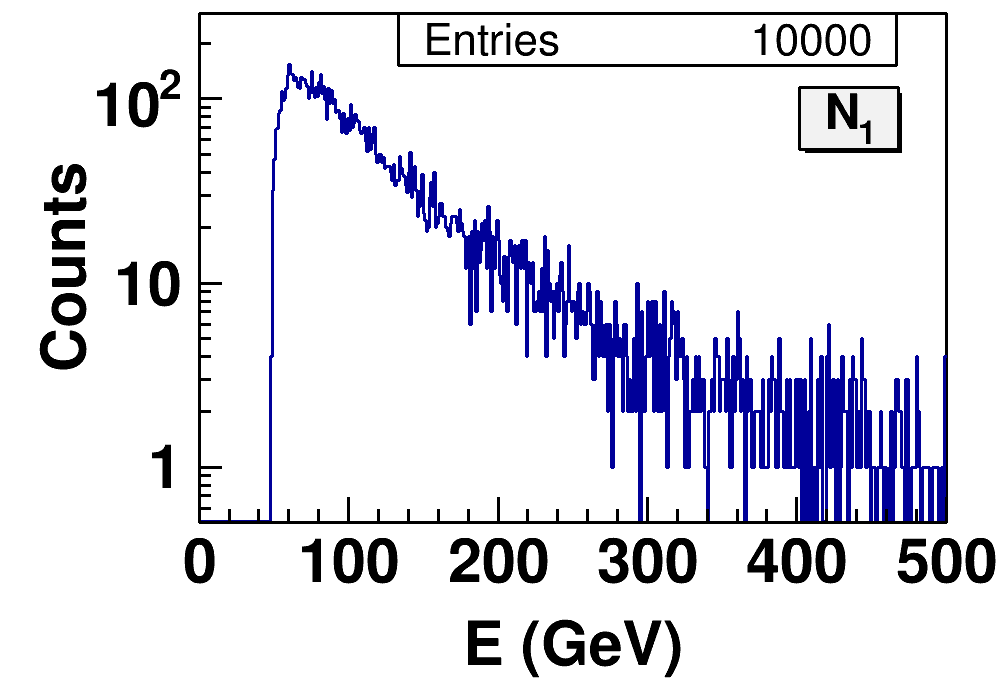} &
    \includegraphics[width=45mm]{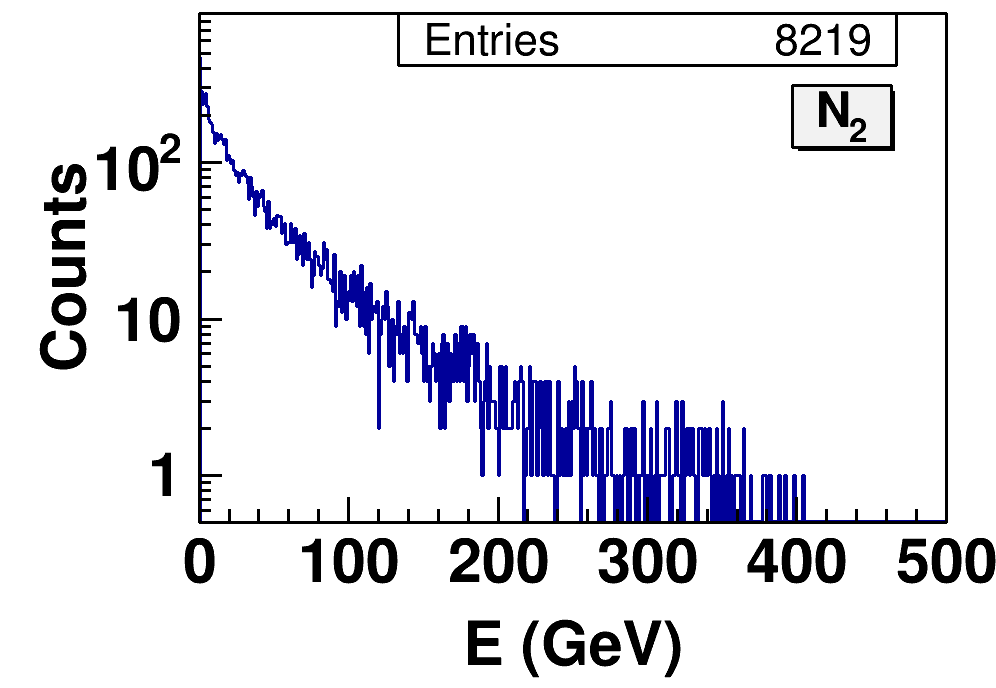}\\
    \includegraphics[width=45mm]{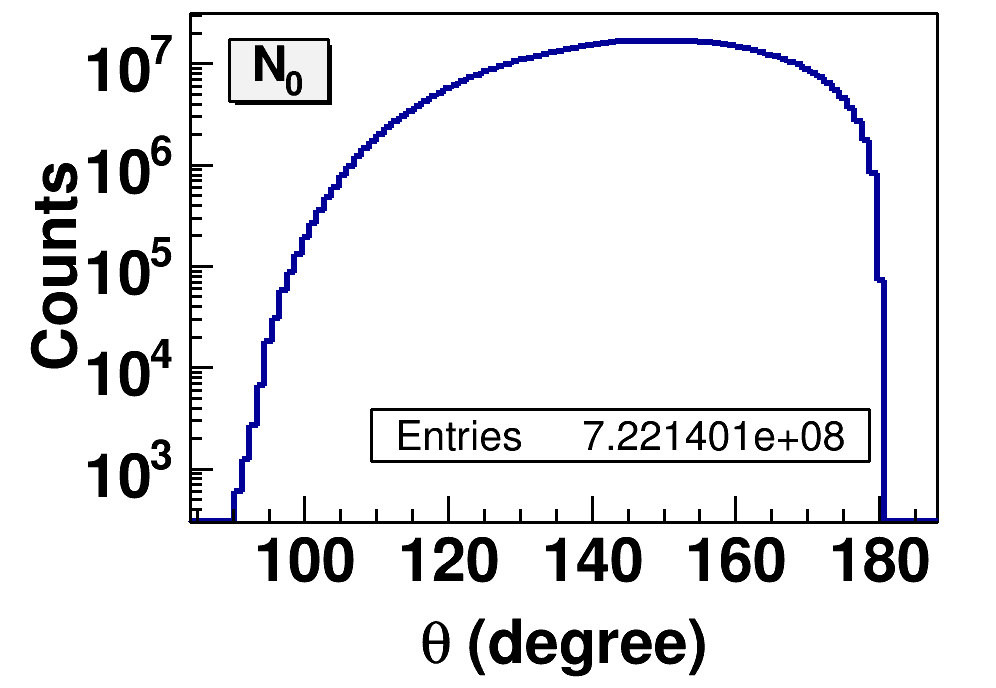} &
    \includegraphics[width=45mm]{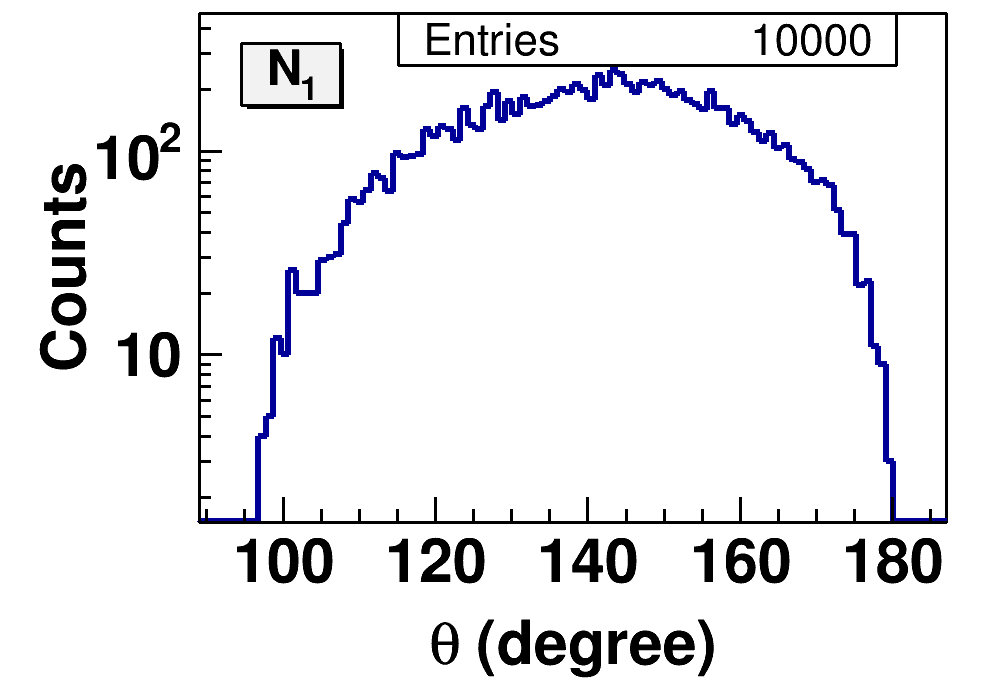} &
    \includegraphics[width=45mm]{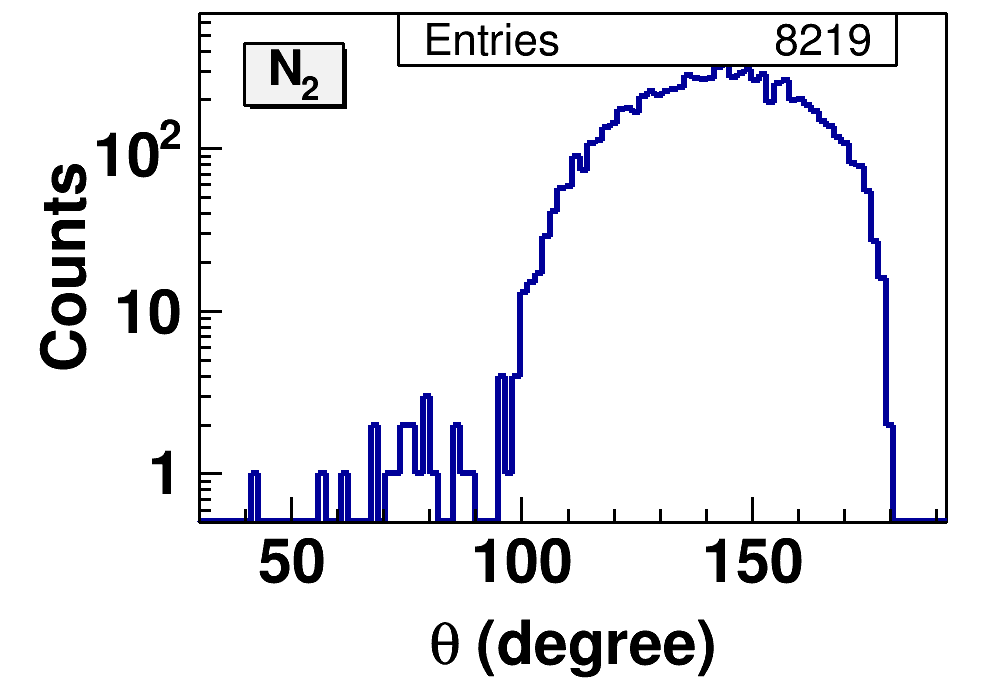} \\
    \includegraphics[width=45mm]{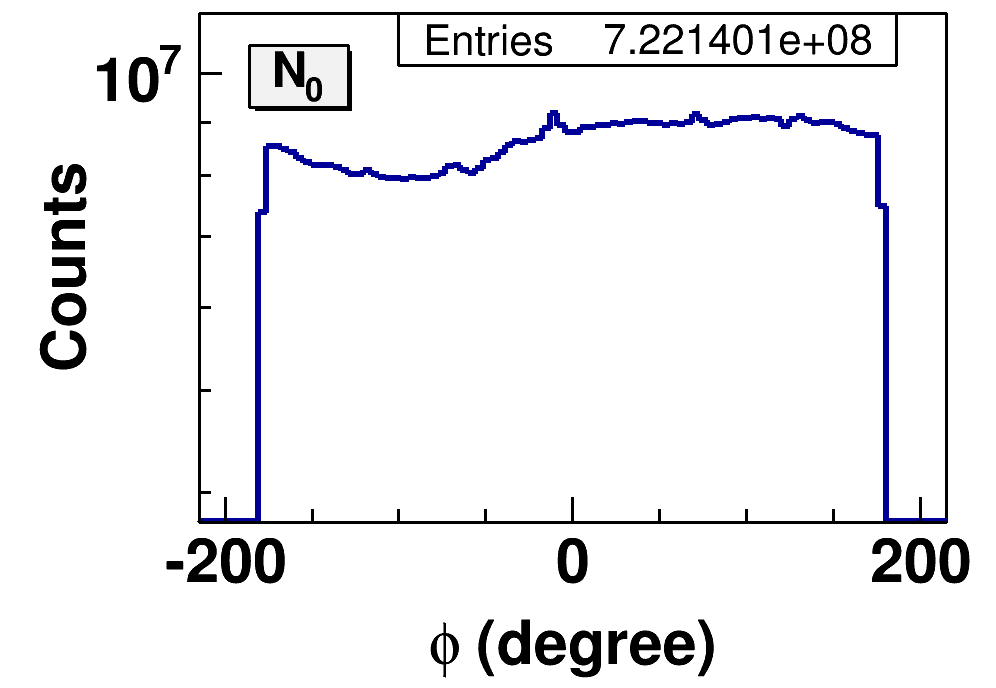}&
    \includegraphics[width=45mm]{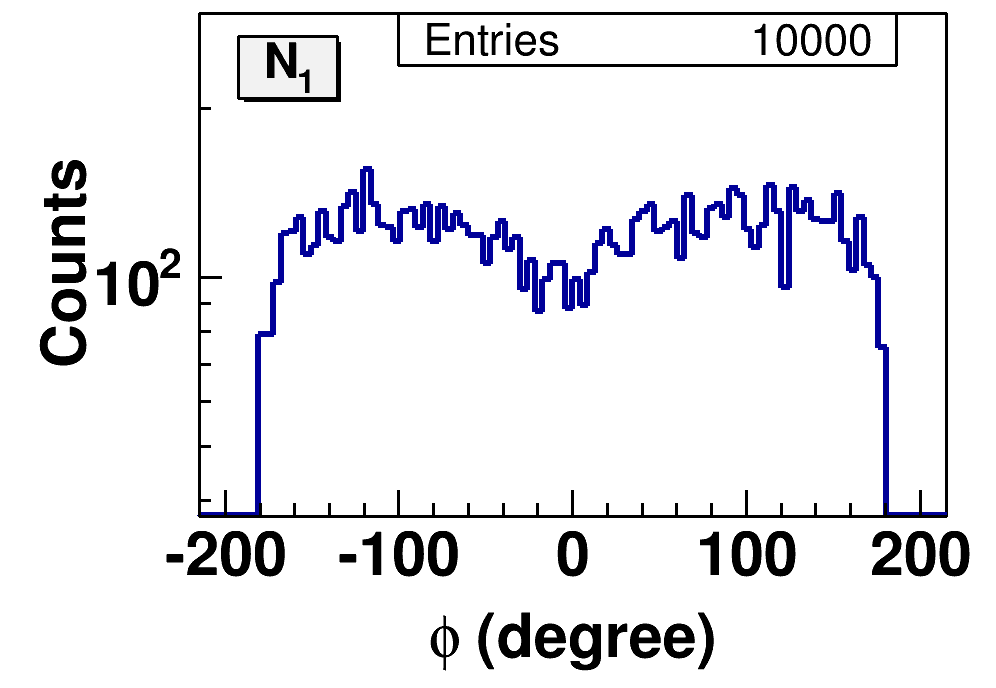}&
    \includegraphics[width=45mm]{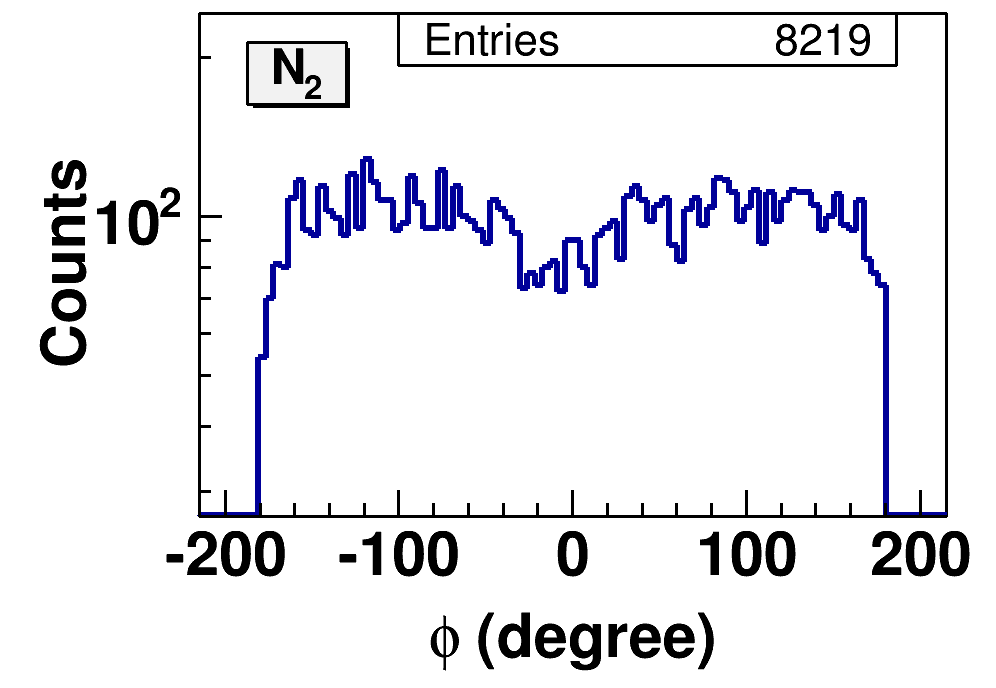}\\
  \end{tabular}
 \caption {\label{fig:E_theta_phi_N0N1N2} $E$ (\textit{top}), $\theta$ (\textit{center}) and $\phi$ (\textit{bottom}) distribution for $N_0$ (\textit{left}), $N_1$ (\textit{center}) and $N_2$ (\textit{right}) for event selections $N_0$, $N_1$, $N_2$ as described in the text.}
\end{figure}

A schematic of the geometry used for the first part of the simulation is shown in Fig.~\ref{fig:schem_part1}.
%%%%%%%%%%%%%%%%%%%%%%%%%%%%%%%%%%%%%%%%%%%%%%%%%%%%%%%%%%%%
%%%%%%%%%%%%%%%%%%%%%%%%%%%%%%%%%%%%%%%%%%%%%%%%%%%%%%%%%%%
In this part, $N_0~(6.48 \times 10^{13})$ cosmic muons of energy E and polar angles ($\theta, \phi$) are generated uniformly over the surface of the rock of area 2\,km $\times$ 2\,km at the sea level using CORSIKA~\citep{corsika} software with the sibyll model~\citep{sybill}. Ideally, a rock with infinite surface area would be required to generate the full cosmic muon spectrum for all possible angles. In the simulation, the area of the 100\,m thick rock is chosen to have transverse dimensions of 2\,km $\times$ 2\,km which mimics an infinite plane as compared to the ICAL surface (48\,m $\times$ 16\,m) and covers $\sim$\,99.9\% of the complete cosmic muon zenith angle spectrum. Muons having low energy would not be able to come out of the rock due to ionization energy loss. Furthermore, muons having an initial direction that doesn't intersect with any side planes of the `volume V', a cuboid of size $\rm(100\,m\,\times\,40\,m\,\times\,40\,m)$ (see Fig.~\ref{fig:schem_part1}), are not expected to contribute to the muon-induced neutral events at the ICAL. The dimension of `volume V' is chosen to be more than the CMVD to take into consideration the change in the direction of muons due to multiple scattering in the rock. Hence, to save computation time, only $N_1 (9\times10^8)$ muons out of $N_0$, having energy $E>\frac{E_{th}}{cos\theta}$ ($E_{th} = 48$\,GeV) and direction ($\theta, \phi$) which intersects any two planes of `volume V' are selected and allowed to propagate through the 100\,m of rock. The ratio between $N_0$ and $N_1$ obtained from the simulation is $7.2\times10^{4}$. This involves a factor due to the solid angle coverage of volume V at the surface of the rock of $6\times10^2$ and a factor of $1.2\times10^2$ from the energy loss of muons in 100\,m of rock. It should be mentioned here that, if a stopping power of $\frac{1}{\rho}\frac{dE}{dx} = 2\,\rm MeVg^{-1}cm^{2}$ is assumed for a MIP (minimum ionizing particle), a muon loses 46\,GeV energy after traversing 100\,m of rock. However, non-MIP muons lose more than $2 \rm MeVg^{-1}cm^{2}$ as they propagate, hence, only $N_2$ muons out of $N_1$ come out of the rock. It is observed that, $N_2$ is $\sim 0.65 N_1$ for $E_{th} = 46$\,GeV and $\sim 0.82 N_1$ for $E_{th} = 48$\,GeV. We have used $E_{th} = 48$\,GeV in this simulation. Throughout the simulation, $\rm SiO_2$ of density 2.32 $\rm g cm^{-3}$ is used as a rock material. The muon-nuclear interaction is not important in this part of the simulation, hence, it was excluded in the physics list of GEANT4. The (x,y) position at the rock surface for $N_0$, $N_1$ and $N_2$ is shown in Fig.~\ref{fig:xy_pos_N0N1N2} and the corresponding $E, \theta$ and $\phi$ distribution is shown in Fig.~\ref{fig:E_theta_phi_N0N1N2} for a subset of the data.
%\footnote{The estimated factor of reduction in $N_0$ from the solid angle coverage of volume V at the surface of the rock is $6\times10^2$ and the same from the energy loss of muons in 100 m of rock is $1.2\times10^2$.}
%%%%%%%%%%%%%%%%%%%%%%%%%%%%%%%%%%%%%%%%%%%%%%%%%%%%%%%%%%%%
%%%%%%%%%%%%%%%%%%%%%%%%%%%%%%%%%%%%%%%%%%%%%%%%%%%%%%%%%%%
%%%%%%%%%%%%%%%%%%%%%%%%%%%%%%%%%%%%%%%%%%%%%%%%%%%%%%%%%%%

\begin{figure*}[ht]
$$\begin{array}{cc}
    \includegraphics[width=0.25\textwidth]{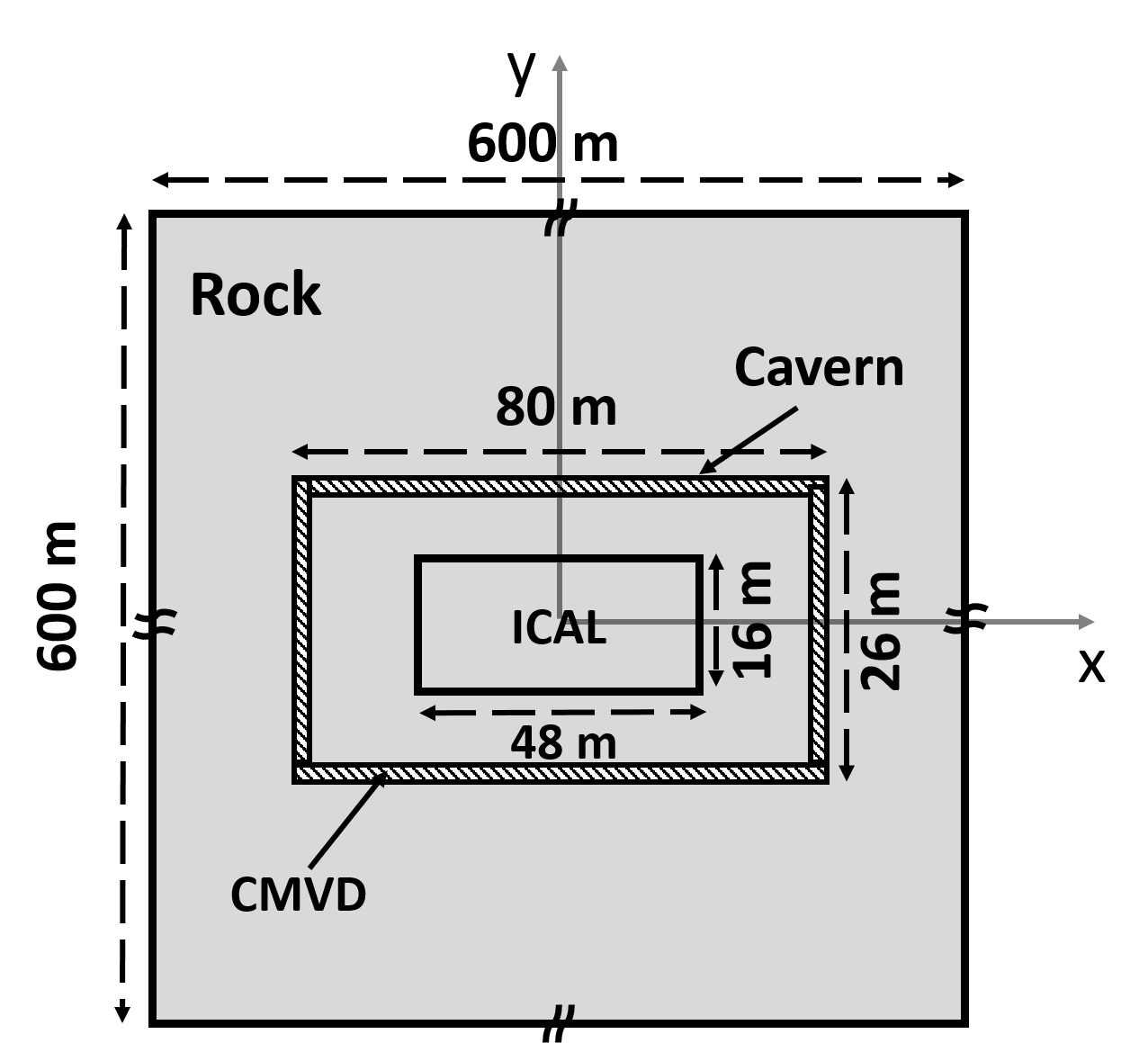} & 
    \includegraphics[width=0.6\textwidth]{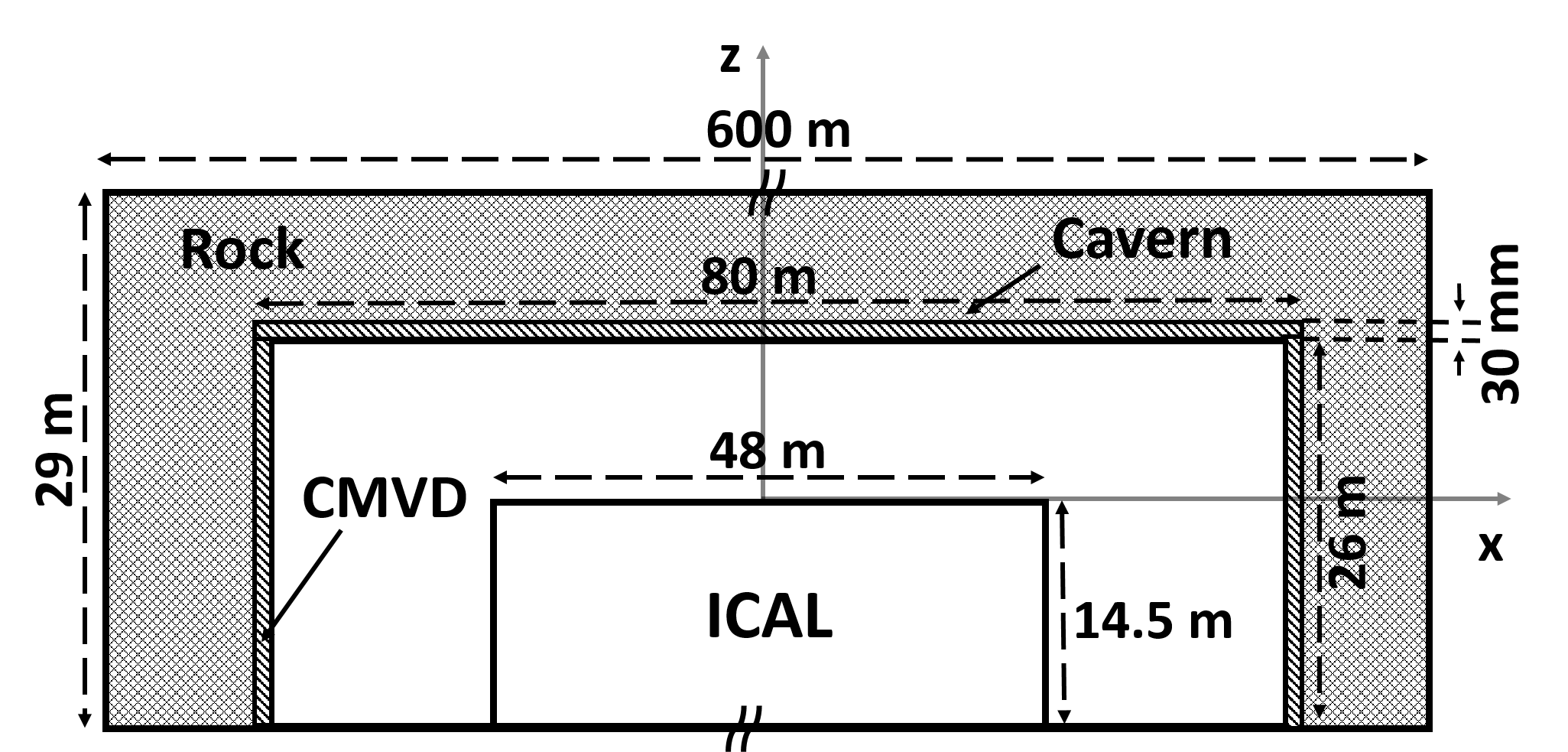}\\ 
    \multicolumn{2}{c}{\includegraphics[width=0.5\textwidth]{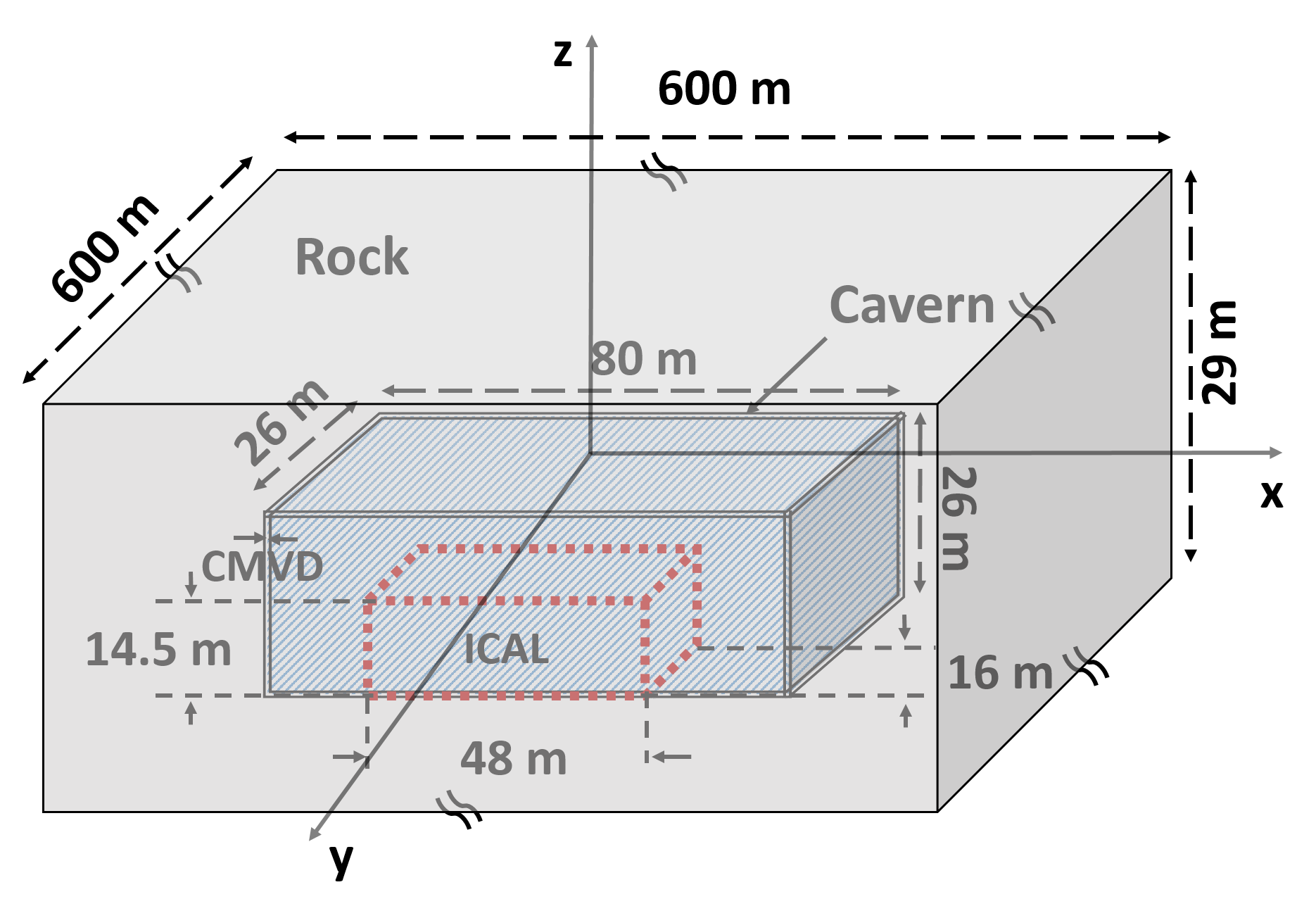}}
\end{array}$$
\caption{A schematic representation for the detector geometry used in the second part of the simulation.}
\label{fig:schem_part2}
\end{figure*}

The detector geometry used for the second part of the simulation is shown in Fig.~\ref{fig:schem_part2}.
%%%%%%%%%%%%%%%%%%%%%%%%%%%%%%%%%%%%%%%%%%%%%%%%%%%%%%%%%%%%
%%%%%%%%%%%%%%%%%%%%%%%%%%%%%%%%%%%%%%%%%%%%%%%%%%%%%%%%%%%
In this part, the muon-induced neutral background for the SICAL detector is studied. For this, $N_2(7.38\times 10^8)$ muons with $E, \theta$ and $\phi$ distribution as shown in Fig.~\ref{fig:E_theta_phi_N0N1N2} (\textit{bottom}) are propagated from the top surface of the 3\,m rock having the same (x,y) coordinates as shown in Fig.~\ref{fig:xy_pos_N0N1N2} (\textit{right}) obtained from the first part. It is legitimate to consider that the neutrals produced in muon-nuclear interactions, mostly from the last part of the 3\,m depth of rock, could exit the rock. The choice of 3\,m was guided by the hadronic interaction length $\lambda$ for rock which is 36\,cm \citep{PDG} i.e. $\sim$10 times smaller. This was verified by performing the simulation for 5\,m and 10\,m of rock which produced, within error, the same number of outgoing neutral particles as with 3\,m rock. Following this argument, only $N_3 (3.69 \times 10^8)$ muons out of $N_2$, that are expected to pass through rock material of 3 m surrounding the cavern, are allowed to propagate. The Kokoulin model \citep{K_model} is used to simulate muon-nuclear interactions. The cross-section for this process was increased by a factor 100 to reduce the computation time. The hadronic interactions of the secondaries are also considered. All the particles (both neutral and charged) that are coming out of the rock and entering through the cavern are recorded in the scintillator of the CMVD. As the muon-nuclear cross-section is increased by a factor of 100, the charged particles (predominantly muons) will create extraneous interactions. Hence, they were not propagated beyond the CMVD once detected in the scintillator. In contrast, the neutrals were allowed to propagate through the cavern towards the ICAL detector.

%\section{False positive event rate in ICAL due to muon-induced neutrals}
\section{Results and discussions}

\begin{figure}[ht]
  \centering
  \begin{tabular}{ccc}
    \includegraphics[width=40mm]{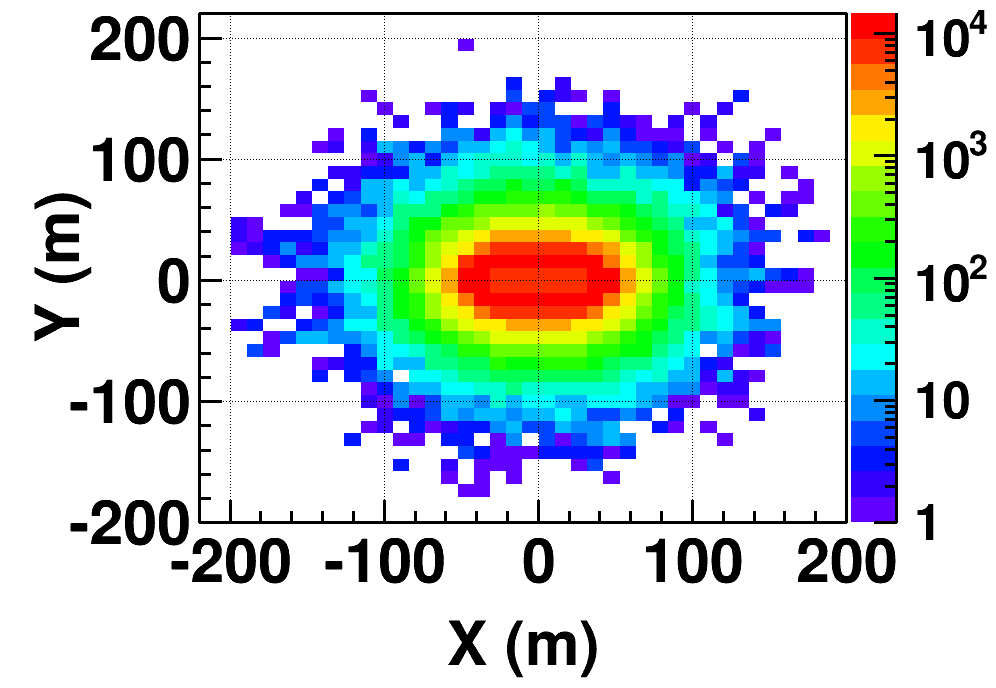} &
    \includegraphics[width=40mm]{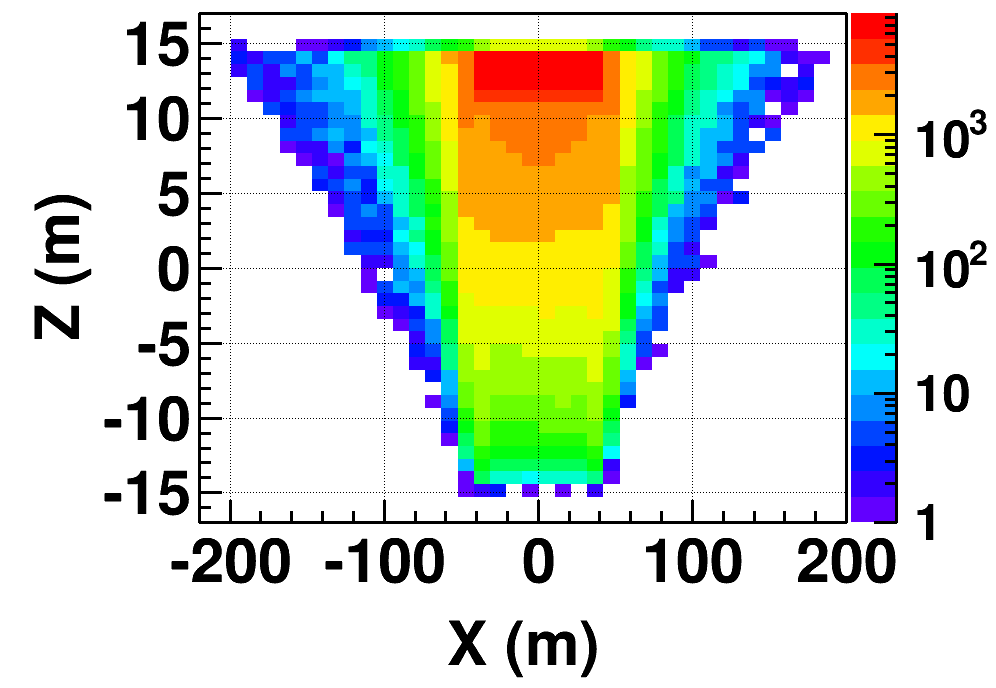}&
    \includegraphics[width=40mm]{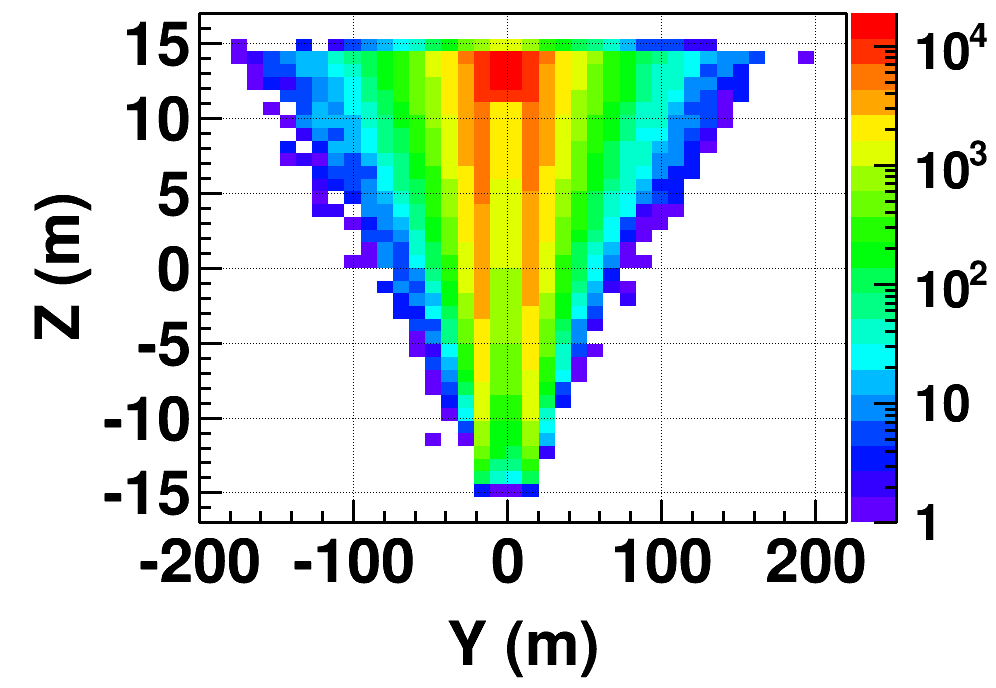}\\
  \end{tabular}
 \caption{The vertices of all the muon-nuclear interaction in different views.}
 
  \label{fig:vertices_intetactn}
\end{figure}

\begin{figure}[ht]
  \centering
  \begin{tabular}{cc}
    \includegraphics[width=60mm]{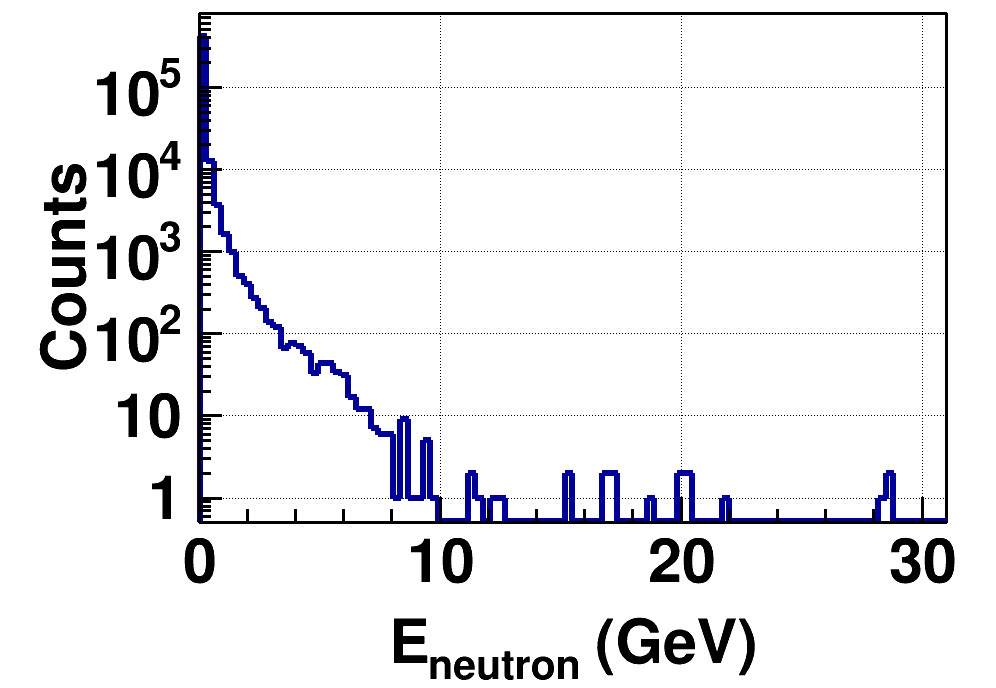} &
    \includegraphics[width=60mm]{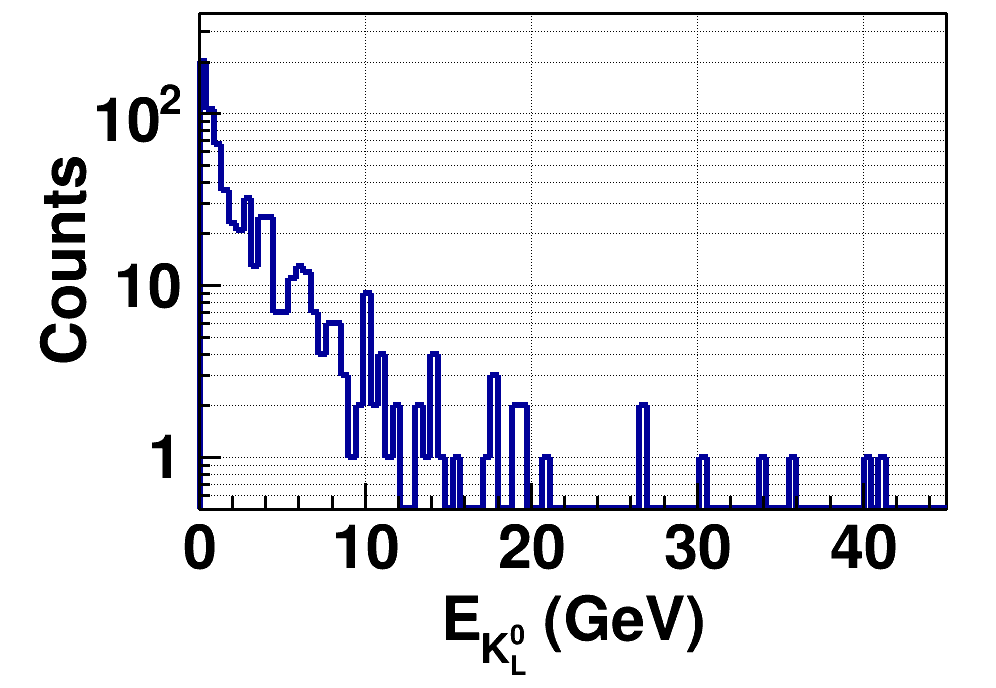}\\
    \includegraphics[width=60mm]{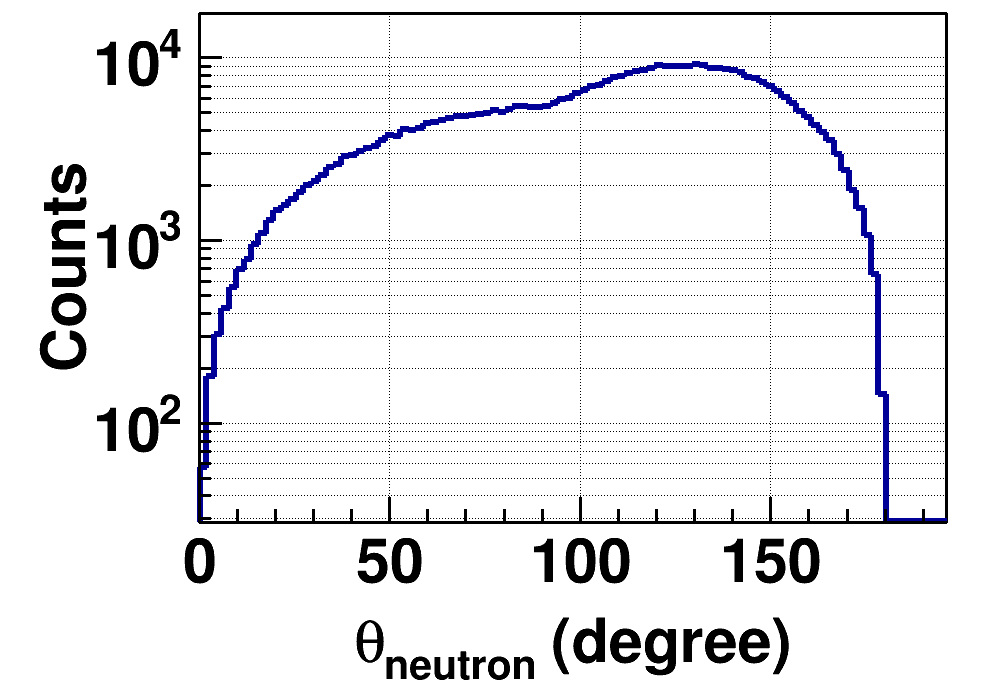}&
    \includegraphics[width=60mm]{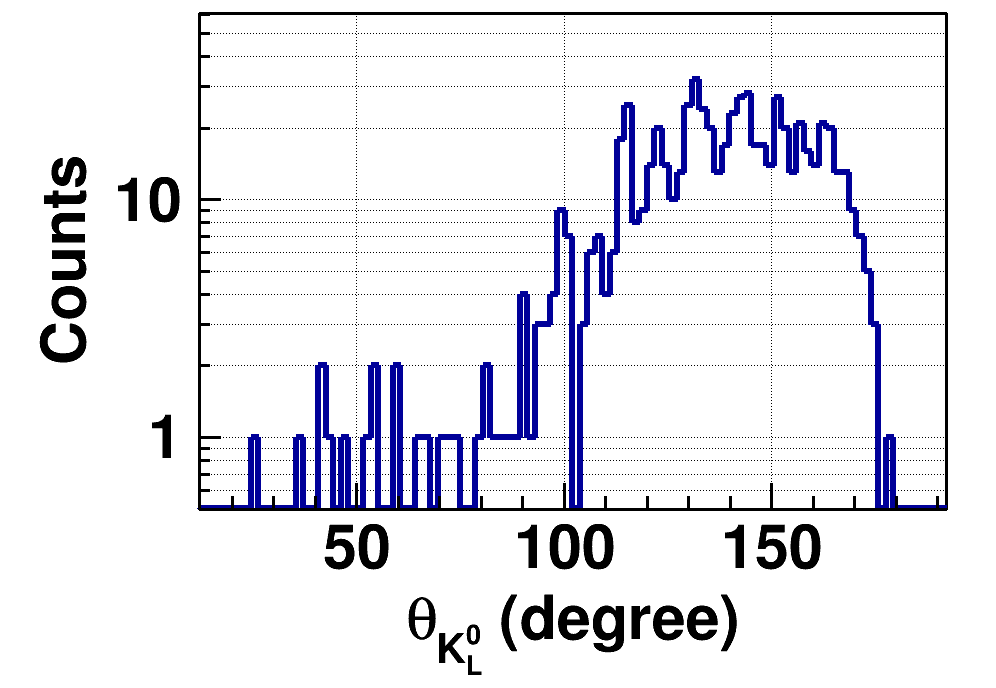}\\
  \end{tabular}
 \caption {\label{fig:muNuc_E_theta_phi} Energy and theta spectra of neutrons  (left top and left bottom) and $K^{0}_{L}$s (right top and right bottom).}
\end{figure}

\begin{figure}[ht]
  \centering
  \begin{tabular}{rr}
    \includegraphics[width=150mm]{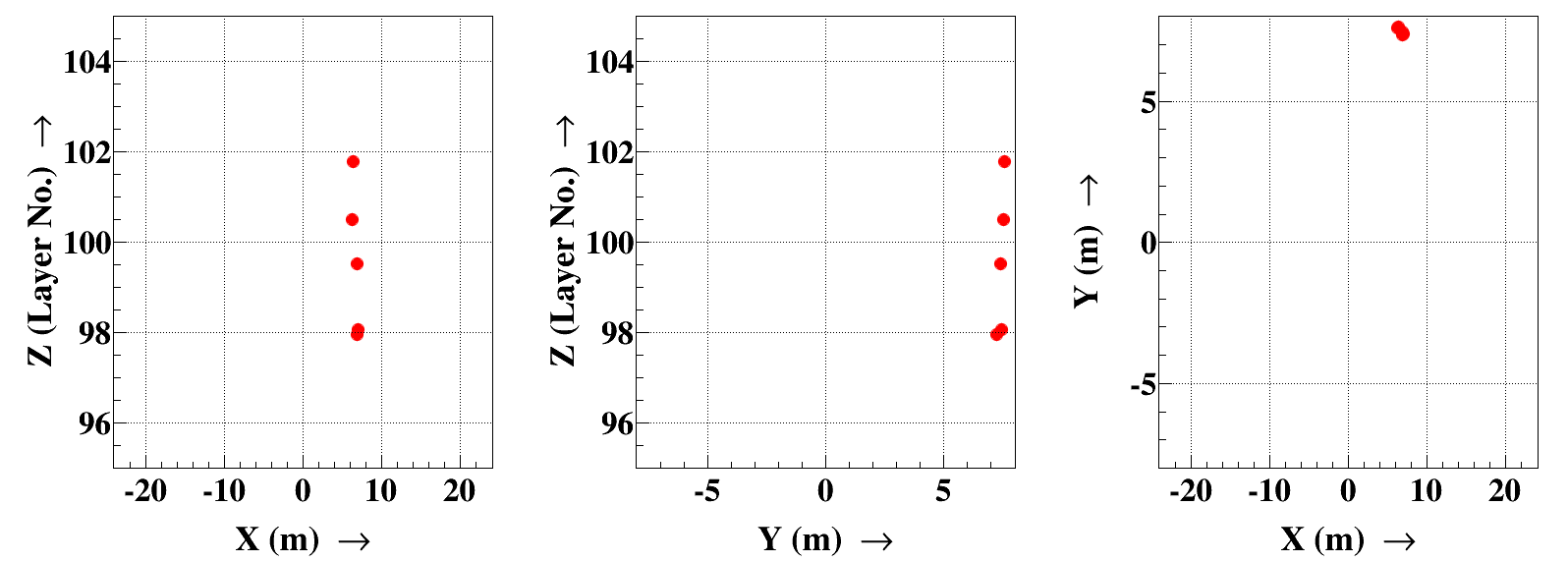}\\
    \includegraphics[width=150mm]{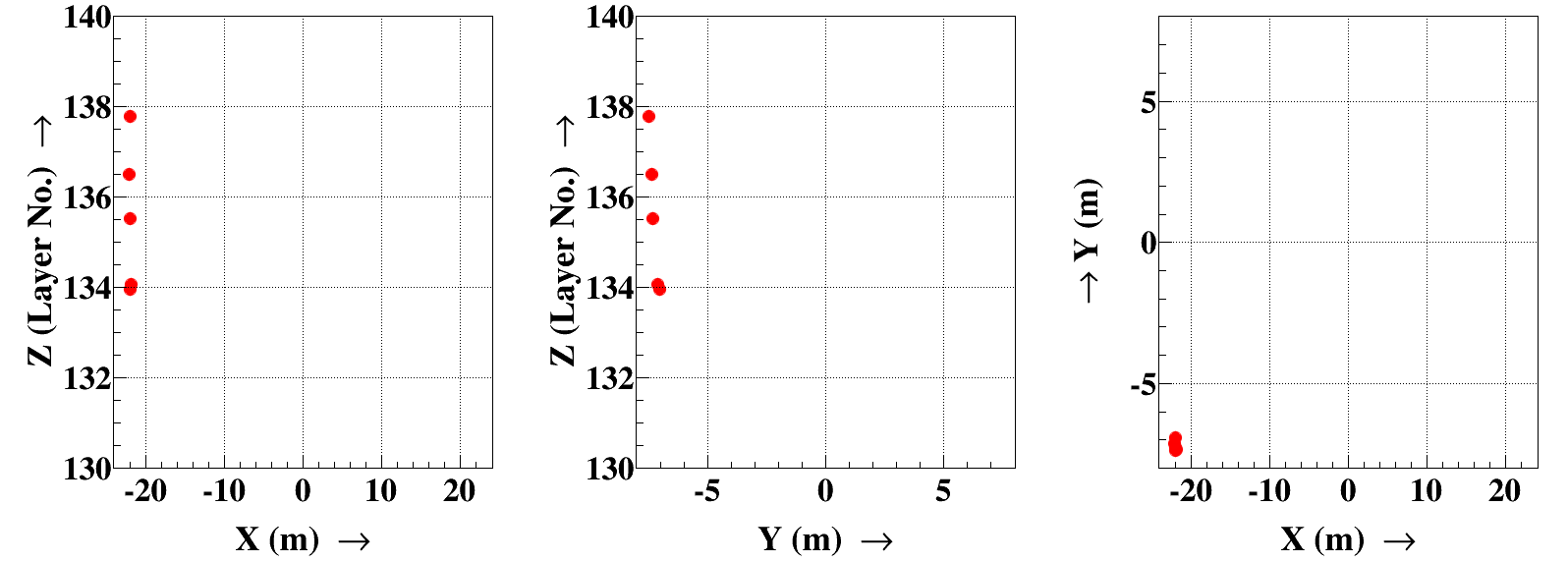}
     \end{tabular}
 \caption {\label{fig:tracks} Reconstructed tracks of the two false positive events in the ICAL detector obtained from the simulation. }
\end{figure}

The total number of muon-nuclear interactions in the rock are found to be $\,5.58\, \times$\,10$^{8}$ and the vertices of these interactions are shown in Fig.~\ref{fig:vertices_intetactn}.%%%%%%%%%%%%%%%%%%%%%%%%%%%%%%%%%%%%%%%%%%%%%%%%%%%%%%%%%%%%
%%%%%%%%%%%%%%%%%%%%%%%%%%%%%%%%%%%%%%%%%%%%%%%%%%%%%%%%%%%
It can be seen that the number of interactions increases with the distance traversed by the muon in the rock as expected. The number of secondaries produced due to muon-nuclear interactions is 4.95\,$\times$\,10$^{9}$ and out of this 2.7\,$\times$\,10$^{9}$ could come out of the rock. Although a neutron cannot produce a muon through its decay, it can interact with the Fe in the ICAL and produce charged pions and kaons which can decay producing muons. The K$^{0}_{L}$ has a decay mode leading to muons and kaons with a branching ratio of 27 \% which makes it important for the present study. The $\pi^{0}$s produced are not of major concern as they have a very short lifetime ($\sim$\,10$^{-16}$ sec) and decay into 2 $\gamma$-rays. Each $\gamma$-ray leads to an electromagnetic shower in the ICAL which can be very well distinguished from the muon trajectories. Consequently, the most important muon-induced neutral background are neutrons and K$^{0}_{L}$. The total number of neutrons and K$^{0}_{L}$ that are produced due to muon-nuclear interaction is $1.08\times 10^9$ and $4.23\times 10^6$, respectively. A typical ($E,\theta$) distribution for the neutrons and the $K^0_L$ at the entrance of the CMVD is shown in Fig.~\ref{fig:muNuc_E_theta_phi}.
%%%%%%%%%%%%%%%%%%%%%%%%%%%%%%%%%%%%%%%%%%%%%%%%%%%%%%%%%%%
%%%%%%%%%%%%%%%%%%%%%%%%%%%%%%%%%%%%%%%%%%%%%%%%%%%%%%%%%%%
It should be pointed out that, the neutral particles should have more than 1\,GeV of energy to produce a charged particle in nuclear interaction, which can pass through five layers of RPC detector in the ICAL. Hence, only those events in which neutrals have energy $>$1\,GeV are considered for further estimation of muon-induced neutral background in the ICAL. The number of such events relevant for this study are\\
a) Events in which neutrals are accompanied by no charged particle\\
b) Events in which neutrals are accompanied by charged particles having kinetic energy less than 10\,MeV as below this energy the charged particle may not give any signal in the CMVD.\\
These events, that are generated as a result of interaction of particles different from a neutrino but are likely to be classified as neutrino induced events in the ICAL detector are called false positive signals. All these events are then reconstructed in the ICAL detector using the Kalman filter technique~\citep{kolahal_kalman,ical_update}. The trajectory of a charged particle due to a false positive signal is considered to be $\nu$-induced muon signal, if it has hits in a minimum 5 layers with $\chi^{2}/ndf < 10$ and is contained within the fiducial volume of the ICAL detector, where the fiducial volume excludes the region of the top 4 layers and 30\,cm from all the four sides of the ICAL detector. Consequently, 2 out of $9\times10^8$ simulated events have resulted in false positive events. These are shown in Fig.~\ref{fig:tracks} and the relevant parameters are listed in Table~\ref{Tab:table1}. This also provides an upper bound of 6.3 false positive events at ICAL with a confidence level~\cite{PDG} (C.L.) of 95\%. Due to almost 100\% efficiency of the veto detector, a large fraction of all the primary muons coming out of the rock would be vetoed. From an earlier measurement with a small Cosmic Muon Veto detector \citep{npanchal} the veto efficiency achieved was 99.987\% which is equivalent to a reduction in muon flux by about 10$^{4}$. Nevertheless, due to the small inefficiency, a part of the total primary cosmic muons will leak through the veto detector undetected. It should be emphasized that, the number of such muons will be comparable to the muon background level in the ICAL detector placed at a depth of 1\,km. This residual primary muon background will be identified and removed in the same way as in the original plan of the ICAL detector with about 1\,km rock over-burden by using the algorithm to detect events in the fiducial volume of the ICAL detector.
%%%%%%%%%%%%%%%%%%%%%%%%%%%%%%%%%%%%%%%%%%%%%%%%%%%%%%%%%%%%
%\begin{figure}[ht]
%  \centering
%    \includegraphics[width=120mm]{name.png} 
% \caption{Reconstructed tracks for a few false positive signals.}
% 
%  \label{fig:tracks}
%\end{figure}

%%%%%%%%%%%%%%%%%%%%%%%%%%%%%%%%%%%%%%%%%%%%%%%%%%%%%%%%%%%

\begin{table}[htbp]\label{Tab:table1}
\centering
\caption{\label{table:tab_4}{Characteristics of the two events that are false positive signal at SICAL. $V_{x}$ and $V_{y}$ denotes the x and y position of the vertices of the reconstructed track.}}
\label{Tab:table1}
\smallskip
\begin{tabular}{|c|c|c|c|c|c|c|}
\hline
No. of hits & $\chi^{2}/ndf$ & $\theta$ (rad) & $\phi$ (rad) & $V_{x}$ (m) & $V_{y}$ (m) & Layer No.\\
\hline
5 & 0.37 & 2.34 & 2.33  & 6.63 & 7.65 & 101\\
5 & 0.24 & 2.17 & -1.44 & -21.8 & -7.48 & 137\\
\hline
\end{tabular}
\end{table}

\section{Estimation of false positive event rate in the SICAL due to muon-induced neutrals.}

In the present study, the muon-induced neutral background at the SICAL is simulated for $N_0 = 6.48 \times 10^{13}$ incident muons at sea level. As mentioned earlier, the cross-section of muon-nuclear interaction is increased by factor of 100 in the simulation in the last 3 m of rock. Therefore, $N_0$ would correspond to an effective number of muons at the sea level $N = 6.48 \times$ 10$^{15}$. The primary cosmic ray muon flux at the sea level is 70\,m$^{-2}$\,sec$^{-1}$\,sr$^{-1}$ \citep{PDG} which integrates to $\sim$\,4.8$\times$10$^{13}$\,/day for a surface area of 2\,km $\times$ 2\,km. Consequently, $N$ muons amounts to a simulation of 135 days. Hence, the false positive rate at the SICAL equates to 0.015\,/day with an upper bound of 0.05\,/day at 95\% confidence level.
The neutrino event rate at INO-ICAL using a Monte Carlo simulation is reported by A. Kumar \textit{et al.}~\citep{white_paper}, where the ICAL geometry and Honda flux \cite{neutrinoflux} was used as inputs in the nuance \citep{nuance} event generator and the neutrino event rate is estimated to be  $\sim$\,4\,/day. The false positive to signal for the SICAL detector is therefore, about $\sim$~0.4\% with an upper bound of $\sim$\,1.2\% at 95\% C.L. which makes the SICAL a feasible proposition. It may be relevant to examine an issue of false vetoes, discussed in the context in the IceCube \citep{yuan}, that may potentially affect the ICAL. The cosmic ray interaction with the upper atmosphere leading to neutrinos is also associated with muons. However, the tracking capability of the ICAL detector in combination with the CMVD should be able to clearly identify the event due to muon neutrinos as opposed to the those originating from the upper atmosphere and giving a track at a physically different location.  
%The neutrino event rate for the ICAL is $\sim$\,3\,/day as discussed in INO white paper \citep{white_paper} in which Monte Carlo simulation results are presented where INO-ICAL geometry and Honda flux \cite{neutrinoflux} has been used as inputs in the NUANCE \citep{nuance} event generator to generate neutrino event rate for ICAL detector. The false positive to signal for the SICAL detector is therefore, about $\sim$~0.5\,\% with an upper bound of $\sim$\,1.6\,\% at 95\,\%C.L. which makes the SICAL a feasible proposition.
%They performed a full Monte Carlo simulation in which the ICAL geometry and Honda flux \cite{neutrinoflux} were used as inputs in the NUANCE \citep{nuance} event generator.

While the energy and angle dependence of the muon  spectrum is pretty robust there could be some leeway in the deep inelastic partial differential cross sections used in GEANT4. A more foolproof test of this idea would be to place a reasonable sized ICAL prototype detector at a shallow depth of $\sim$\,30\,m, enclose it in a CMVD and quantify the false positive which could mimic neutrino events in the ICAL, at the same time comparing with the simulations at that depth.  

\section{Summary}

In summary, we have presented results of simulations which support the possibility of locating the INO-ICAL detector at a shallow location with a rock overburden of $\sim$\,100\,m when used with an efficient cosmic ray shield with an efficiency of about 99.99\% for detecting charged particles. The main background is due to neutral, long lived and energetic particles produced in the last few metres of rock either unaccompanied by or associated charged particles which go undetected. This fraction has been estimated to be much smaller ($\sim$\,0.4\% with an upper bound of $\sim$\,1.2\% at 95\% C.L.) than the signal due to atmospheric neutrinos. Therefore, the proposal of a the SICAL detector opens up the possibility of having a much larger choice of locations, saving time due to the shorter tunnel and allowing for much larger caverns. However, it must be mentioned that it is perhaps necessary, and prudent that a proof-of-principle detector be built at a shallow depth, perhaps even at 30\,m rock overburden, together with the Cosmic Veto Detector to validate the simulation.

\acknowledgments

We would like thank M.V.N. Murthy, Naba K. Mondal, D. Indumathi, Paschal Coyle, John G. Learned, Morihiro Honda, Thomas K. Gaisser, Shashi R. Dugad, Deepak Samuel and Satyajit Saha for useful comments and suggestions and Apoorva Bhatt and P. Nagaraj for help in simulation related issues. Finally, we would also like to thank the anonymous referee for critical but useful comments.

% Please avoid comments such as "For a review'', "For some examples",
% "and references therein" or move them in the text. In general,
% please leave only references in the bibliography and move all
% accessory text in footnotes.

% Also, please have only one work for each \bibitem.


\begin{thebibliography}{99}

\bibitem{akumar}
M.S. Athar et al., \emph{INO Collaboration:Project Report}, {\bf Vol. I} (2006).

\bibitem{ghosh2013}
A. Ghosh et al., \emph{Determining the neutrino mass hierarchy with INO, T2K, NOvA and reactor experiments}, \emph{Journal of High Energy Physics} {\bf 2013} (2013) 9.

\bibitem{npanchal}
N. Panchal et al., \emph{A compact cosmic muon veto detector and possible use with the Iron Calorimeter detector for neutrinos}, \emph{Journal of Instrumentation} {\bf 12} (2017) T11002.

\bibitem{muSR}
Neha et al., \emph{Can Stopped Cosmic Muons Be Used to Estimate the Magnetic Field in the Prototype ICAL Detector?}, \emph{XXII DAE High Energy Physics Symposium-Proceedings, Springer Science and Business Media, LLC} {\bf 203} (2018) 93.

\bibitem{dash1}
N. Dash, V.M. Datar and G. Majumder, \emph{Sensitivity for detection of Dark Matter particle using ICAL at INO}, \emph{Pramana} {\bf 86} (2016) 927.

\bibitem{dash2}
N. Dash, V.M. Datar and G. Majumder, \emph{Sensitivity of INO-ICAL detector to Magnetic Monopoles}, \emph{Astropart. Phys.} {\bf 70} (2015) 33.

\bibitem{ical_update}
S. Seth, A. Bhatt, G. Majumder and A. Mishra \emph{Update of INO-ICAL reconstruction algorithm} \emph{Journal of Instrumentation} {\bf 13} (2018) P09015. 

\bibitem{GEANT4} 
GEANT4 collaboration, S. Agostinelli et al., \emph{GEANT4: A Simulation toolkit}, \emph{Nucl. Instrum. Meth.} A {\bf 506} (2003) 250.

\bibitem{corsika} 
D. Heck, J. Knapp, J.N. Capdevielle, G. Schatz and T. Thouw, \emph{CORSIKA: A Monte Carlo Code to Simulate Extensive Air Showers}, \emph{Forschungszentrum Karlsruhe Report FZKA} {\bf 6019} (1998).

\bibitem{sybill}
Fletcher, R. S. and Gaisser, T. K. and Lipari et al. \emph{sibyll: An event generator for simulation of high energy cosmic ray cascades}, \emph{Phys. Rev. D} {\bf 50} (1994) 5710-5731.

\bibitem{PDG}
C. Patrignani et al. (Particle Data Group), \emph{Chin. Phys. C} {\bf 40} (2016) 100001.

\bibitem{K_model}
A. G. Bogdanov et al., \emph{Geant4 simulation of production and interaction of muons} \emph{IEEE Trans. on Nuc. Sci.} {\bf 53} (2006) No. 2.

\bibitem{kolahal_kalman} 
Kolahal Bhattacharya, Arnab K. Pal, Gobinda Majumder and Naba K. Mondal, \emph{Error propagation of the track model and track fitting strategy for the Iron CALorimeter detector in India-based neutrino observatory}, \emph{Computer Physics Communications} {\bf 185} (2016) 3259-3268.

\bibitem{white_paper}
Kumar, A. et al., \emph{Invited review: Physics potential of the ICAL detector at the India-based Neutrino Observatory (INO)}, \emph{Pramana - J Phys} {\bf88} (2017) 79.

\bibitem{neutrinoflux}
M. Honda et al., \emph{Atmospheric neutrino flux calculation using the NRLMSISE-00 atmospheric model}, \emph{Phys. Rev. D} {\bf 92} (2015) 023004.

\bibitem{nuance}
D Casper, \emph{The nuance neutrino physics simulation, and the future}, \emph{Nucl. Phys. Proc. Suppl.} {\bf 112} (2002) 161.

\bibitem{yuan} 
M.~G.~Aartsen \textit{et al}., \emph{Measurement of Atmospheric Neutrino Oscillations at 6--56 GeV with IceCube DeepCore} \emph{Phys. Rev. Lett.} {\bf120} (2018) 071801. 

% Please avoid comments such as "For a review'', "For some examples",
% "and references therein" or move them in the text. In general,
% please leave only references in the bibliography and move all
% accessory text in footnotes.

% Also, please have only one work for each \bibitem.


\end{thebibliography}
\end{document}